\documentclass{ifacconf}
\usepackage{amsmath}
\usepackage{amsfonts}
\usepackage{graphicx}      
\usepackage{natbib}        
\usepackage{subcaption}
\usepackage{algorithm}
\usepackage{algpseudocode}  
\usepackage{enumitem}
\usepackage{eso-pic}
\graphicspath{{./figures/}}

\usepackage{pgfplots} 
\usepackage{tikz}
\usepackage{color}
\pgfplotsset{compat=1.14}
\usetikzlibrary{plotmarks}	
\usetikzlibrary{shapes,arrows}
\usetikzlibrary{calc, decorations.pathmorphing, fadings, shadings}
\definecolor{mycolor1}{rgb}{0.00000,0.44700,0.74100}
\definecolor{mycolor2}{rgb}{0.85000,0.32500,0.09800}
\definecolor{mycolor3}{rgb}{0.92900,0.69400,0.12500}
\definecolor{mycolor4}{rgb}{0.49400,0.18400,0.55600}
\definecolor{mycolor5}{rgb}{0.63500,0.07800,0.18400}
\definecolor{mycolor6}{rgb}{0.30100,0.74500,0.93300}

\newcommand{\plotref}[1]{}
\newtheorem{remark}{Remark}

\newcommand{\tikzline}[1]{(\protect\tikz[baseline=-0.6ex,x=1pt,y=1pt]{ \protect\draw[#1,thick] [-] (0,0) -- (10,0);})}
\newcommand{\tikzmarkline}[1]{(\protect\tikz[baseline=-0.6ex,x=1pt,y=1pt]{ \protect\draw[#1,thick] [-] (0,0) -- (10,0);\protect\draw[color=#1, thick] (5,0) circle (2pt)})}
\newcommand{\tikzdashedline}[1]{(\protect\tikz[baseline=-0.6ex,x=1pt,y=1pt]{ \protect\draw[#1,thick,dash pattern=on 2pt off 2pt] [-] (0,0) -- (10,0);})}

\begin{document}

			\AddToShipoutPictureBG*{%
	\AtPageUpperLeft{%
		\setlength\unitlength{1in}%
		\hspace*{\dimexpr0.5\paperwidth\relax}
		\makebox(0,-2)[c]{
			\parbox{\paperwidth}{ \centering 
				Enzo Evers, Temperature-Dependent Modeling of Thermoelectric Elements, \\
				In {\em 21st IFAC World Congress}, Berlin, Germany, 2020 } }%
}}
	
\begin{frontmatter}

\title{Temperature-Dependent Modeling of Thermoelectric Elements} 

\thanks[footnoteinfo]{This work is supported by the Advanced Thermal Control consortium (ATC), and is part of the research programme VIDI with project number 15698, which is (partly) financed by the Netherlands Organization for Scientific Research (NWO).}

\author[First]{Enzo Evers}
\author[First]{Rens Slenders}
\author[Second]{Rob van Gils}
\author[First]{Tom Oomen}

\address[First]{Control Systems Technology, Mechanical Engineering Department, Eindhoven University of Technology, Eindhoven, The Netherlands, 
   (e-mail: E.Evers@tue.nl).}
\address[Second]{Philips Innovation Services, Eindhoven, The Netherlands.}

\begin{abstract} 
Active thermal control is crucial in achieving the required accuracy and throughput in many industrial applications, e.g., in the medical industry, high-power lighting industry, and semiconductor industry. Thermoelectric Modules (TEMs) can be used to both heat and cool, alleviating some of the challenges associated with traditional electric heater based control. However, the dynamic behavior of these modules is non-affine in their inputs and state, complicating their implementation. To facilitate advanced control approaches a high fidelity model is required. In this work an approach is presented that increases the modeling accuracy by incorporating temperature dependent parameters. Using an experimental identification procedure, the parameters are estimated under different operating conditions. The resulting model achieves superior accuracy for a wide range of temperatures, demonstrated using experimental validation measurements.
\end{abstract}


\end{frontmatter}


\section{Introduction}

Advanced thermal control is a crucial area of research and development, especially in the medical, high-power lighting, and semiconductor industry. For example, in the medical field, diagnostic platforms are used to process extremely small fluid volumes (e.g. blood, saliva) \citep{yager_microfluidic_2006}. The temperature of these fluid volumes needs to be accurately controlled. Hand held devices are designed to significantly reduce analysis time and reagent costs \citep{jiang_nonlinear_2011}. Another example is in high-power LED lighting. These LEDs generate significant amounts of heat and should be actively controlled to achieve sufficient light quality and an increased lifespan \citep{kaya_experimental_2014}. Finally, in the semiconductor industry, wafer scanners are used to produce integrated circuits that need to achieve a positioning accuracy of nanometers. Therefore, thermal control is an important aspect in their mechatronic design \citep{bos_io_2018}, since the current performance of these high precision systems is often limited by thermal induced deformations. In \citet{saathof_deformation_2016} selective local heating is employed to control the thermal induced deformations in a mirror system. \\
In these research fields, thermoelectric modules (TEMs) have received increasing attention over traditional water conditioning circuits because they have compact dimensions, have no moving parts, and have active heating and cooling capabilities. Moreover, since these TEMs are not limited to heating, this alleviates some of the challenges \citep{evers_identifying_2019} associated with using heating elements for active thermal control. 

The thermodynamics of Peltier elements are non-affine as function of state and input, complicating the controller design. Standard linear control methods (e.g. PID control) could be unreliable because stability, robustness, and performance cannot be guaranteed. Therefore, nonlinear control methods can be considered. In recent literature several methods have been developed. In \citet{shao_lpv_2014}, a linear-parameter-varying approach is used to control the nonlinear system, which linearizes the nonlinear system at different operating points. For each operating point a different controller is synthesized. In \citet{guiatni_thermoelectric_2007}, a sliding- mode controller is used, which applies a state feedback. The state feedback ensures that all trajectories move towards a stable sliding manifold. Lastly, in \citet{bos_io_2018} and \citet{van_gils_practical_2017}, the nonlinear system is partially linearized using a feedback linearization by creating a new virtual input that has linear input-to-output (IO) dynamics. This facilitates the use of conventional linear control approaches.

In the work by \citet{van_gils_practical_2017}, the cold side of a TEM is thermally controlled for a large temperature range from 5 to 80 $^\circ C$. However, the feedback linearization yields some residual nonlinear dynamics. Moreover, in \citet{bos_io_2018,van_gils_practical_2017} a nonlinear observer design is recommended since it is often impractical to install temperature sensors around the point of interest (POI). For example, in a diagnostic platform fluid temperature must be accurately controlled but placing a temperature sensor in the fluid is undesired due to hygiene constraints. 

In view of control and to facilitate the implementation of both accurate linearization methods and observer design, a high-fidelity model of the TEM is required. In literature, often a limited operating temperature for the TEM is considered, allowing the model to be simplified by using temperature independent parameters.
In this paper a significantly larger operating temperature range is considered, e.g., from 5 to 80 $^\circ C$, necessitating the inclusion of temperature dependency in the simulation model. 
This paper expands on previous results in literature \citep{mitrani_methodology_2004} and illustrates the effectiveness of dedicated identification experiments. 
The main contributions of this paper are:
\begin{itemize}
	\item[C1] Incorporating temperature dependent parameters in the thermodynamical TEM model.
	\item[C2] A suitable identification procedure to determine the parameters over a wide temperature range.
	\item[C3] Experimental identification and model validation using a dedicated experimental setup. 
\end{itemize}

\begin{figure}
	\centering
	\begin{subfigure}[t]{\linewidth}
		\centering
		\setlength{\fboxsep}{0pt}
		\fbox{\includegraphics[width=0.99\linewidth]{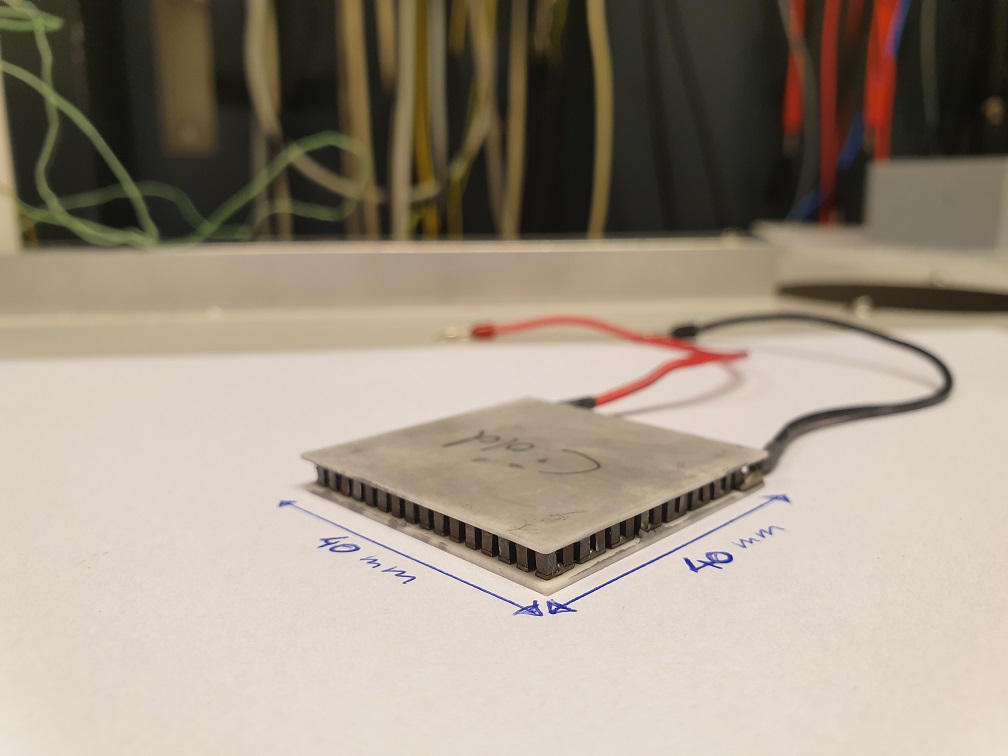}}
		\caption{Photograph.}
		\label{fig:img20191105153224-1}
	\end{subfigure} \\
	\begin{subfigure}[t]{\linewidth}
		\centering
		\includegraphics[width=0.99\linewidth]{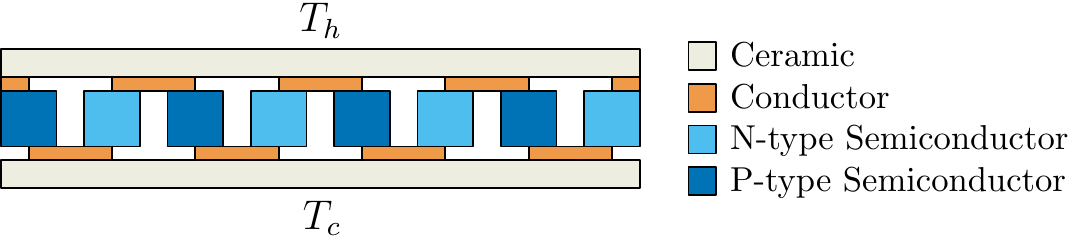}
		\caption{Schematic representation.}
		\label{fig:peltier}
	\end{subfigure}
	\caption{Photograph and schematic representation of the thermoelectric module used in this paper. The semiconductor elements are contained between two ceramic plates.}
	\label{fig:peltier_pic_and_fig}
\end{figure}
\section{TEM Modeling}\label{sec:3}
In this section the first principle model describing a thermoelectric module is derived. Emphasis is placed on temperature dependent modeling, and it is shown that including this dependency can increase modeling accuracy for a wide temperature range.
\subsection{First principles}
A common approach \citep{lineykin_modeling_2007,van_gils_practical_2017,bos_io_2018,fraisse_comparison_2013} to modeling a thermoelectric module is lumped-capacitance discretization, e.g., the module is subdivided into lumps of uniform temperature. A similar approach is taken in this paper by dividing the module, shown in Fig. \ref{fig:peltier_pic_and_fig}, into a hot and cold side where each ceramic plate is a single lump. 

The thermal dynamics of a single TEM are described by including 3 phenomena: 1) The Fourier effect $Q_f$, 2) Joule heating $Q_j$ and 3) the Peltier effect $Q_p$. 
\paragraph*{Fourier effect} The Fourier effect $Q_f$ describes the energy transfer through conduction between the 2 sides of the TEM and it is given by
\begin{equation}
Q^{1\rightarrow 2}_f = \dfrac{1}{R_{1\rightarrow 2}}(T_1-T_2)
\end{equation}
for conduction from temperature $T_1$ to $T_2$ where $\dfrac{1}{R_{1\rightarrow 2}} = \dfrac{k\cdot A}{d}$ with $k$ the conductivity of the material in $\mathrm{W/m\cdot K}$, $A$ the area in $\mathrm{m^2}$ perpendicular to the heat flow and $d$ in $\mathrm{m}$ the distance of the heat flow path.
\paragraph*{Joule heating} Joule heating $Q_j$ occurs when an electrical current flows through a resistive element, in this case the TEM, and is given by 
\begin{equation}
Q_j = R_{m}I^2
\end{equation}
where $R_{m}$ is the electrical resistance in $\mathrm{\Omega}$ of a single TEM and $I$ is the current in $\mathrm{A}$. 
\begin{remark}
	It can be observed that the term $Q_j$ is nonlinear in the input current $I$, complicating the implementation of linear controllers. Several solutions are available, e.g., input-output linearization or nonlinear control design, these are outside the scope of the current work.
\end{remark}
\paragraph*{Peltier effect} The Seebeck effect describes the occurrence of an electrical potential over a semi-conductor in the presence of a temperature gradient. The analogous Peltier effect describes the occurrence of a heat flow over a semi-conductor in the presence of an electrical potential difference and resulting current. While they are manifestation of the same physical phenomena, for the thermal dynamics the latter is described as
\begin{equation}
Q_p = S_mTI
\end{equation}
where $S_m$ is the Seebeck coefficient of the TEM and $T$ is the temperature at the cold/hot side. 

Under the assumption that the Joule heating $Q_j$, that is generated in the semi-conductors, see Fig. \ref{fig:peltier}, is divided equally over the hot and cold side the energy balance for the hot and cold side is given by

\begin{align}
Q_c &= Q^{h\rightarrow c}_f + \frac{1}{2}Q_ j - Q_p + Q^{env\rightarrow c}\\
\nonumber	&= \dfrac{1}{R_{h\rightarrow c}}(T_h-T_c) + \frac{1}{2}R_mI^2 -  S_mT_cI +  Q^{env\rightarrow c} \\
Q_h &= Q^{c\rightarrow h}_f + \frac{1}{2}Q_ j + Q_p + Q^{env\rightarrow h} \\  
\nonumber	&= \dfrac{1}{R_{c\rightarrow h}}(T_c-T_h) + \frac{1}{2}R_mI^2 +  S_mT_hI + Q^{env\rightarrow h} 
\end{align}
where $\dfrac{1}{R_{h\rightarrow c}} = \dfrac{1}{R_{c\rightarrow h}}$ and $Q^{env\rightarrow c,h}$ accounts for any thermal interaction with the environment, indicated by the superscript $^{env}$, i.e., the ambient air or neighboring lumps.

By constructing an energy balance equation for each lump, a complete model can be constructed including the TEM and any connecting elements. This is often done by constructing a state-space model, where the states $x = T_{(1,\dots,N_x)}$, with $N_x$ the number of states, represent the temperature of the lumps with corresponding state equations

\begin{equation}
E_n \dot{x}_n = \sum Q_n, \quad n \in \{1, \dots N_x \}
\end{equation}

where $E_n = m_n c_n$ is the thermal capacitance of the lump $n$ with $m_n$ the mass in $\mathrm{kg}$ and $c_n$ the specific heat capacity in $\mathrm{J/kg K}$. By collecting these differential equations the state-space model of a system is given by
\begin{align}
\dot{x} &= Ax + F^{NL}(x,u) + Bu \label{eq:state_space} \\
\nonumber y &= Cx \\
x(0)_n &= T_a \forall n \in \{1,2,\dots,N_x\}
\end{align}
where $F^{NL}(x,u) $ is a nonlinear function depending on $x$ and the inputs $u$ and $y$ is the output of the system, often corresponding to a temperature, $x(0)_n$ the initial condition of the state and $T_a$ the ambient temperature. By incorporating the nonlinear, e.g., the joule heating $Q_j$ and state-dependent dynamics in $F^{NL}(x,u) $ a full system model is constructed. 
\subsection{Temperature Dependent Modeling} \label{sec3_2}
Employing constant parameters in the model \eqref{eq:state_space} often yields sufficiently accurate results, as demonstrated in \citet{bos_io_2018}, for systems that operate in a limited temperature range. For the systems considered in this paper, e.g., a blood diagnostic device that cycles between $20$ and $80$ degrees Celsius, this is often not sufficient and temperature dependencies must be taken into account.

Including the temperature dependency in \eqref{eq:state_space} is done by modeling the parameters $S_m$ and $R_m$ as a function of the average temperature $T_{avg}$ of the TEM, i.e., $S_m(T_{avg}), R_m(T_{avg})$, where 
\begin{equation}
T_{avg} = \dfrac{T_c + T_h}{2}.
\end{equation}
\begin{remark}
	While the conductivity of the TEM is also considered temperature dependent in some literature. In this paper, this could not be concluded and it is considered outside the scope of the current research. 
\end{remark}
\subsection{Identifying parameters} \label{sec:method_id_param}
\begin{figure}
	\centering
	\includegraphics[width=1\linewidth]{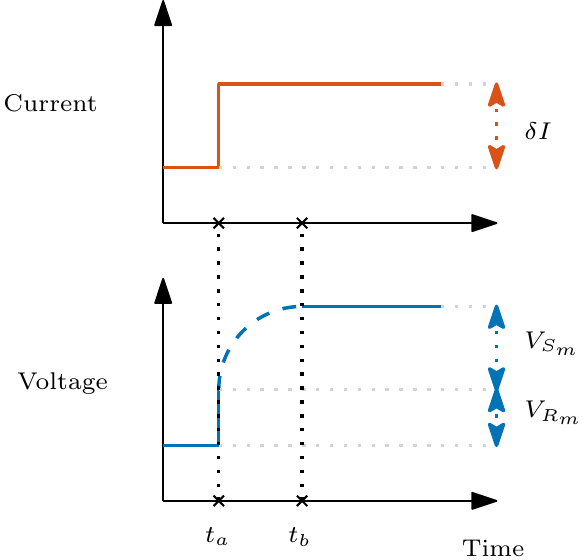}
	\caption{Illustration of the voltage profile following a current step $\delta I$. Since the current amplifier is in closed-loop, the voltage is increased to compensate for the back EMF voltage $V_{S_m}$.}
	\label{fig:peltiervoltage}
\end{figure}

Identifying the electrical resistance $R_m$ and Seebeck coefficient $S_m$ is done by measuring the electrical potential $V_{TEM}$ required to induce a fixed current $I_{TEM}$ in a single TEM. The total voltage is given by
\begin{align}
\nonumber V_{TEM} &= V_{R_m} + V_{S_m} \\ 
&= R_m(T_{avg})I_{TEM} + S_m(T_{avg})(T_h - T_c) \label{eq:tem_voltage}
\end{align}
where $I_{TEM}$ is the current output in $\mathrm{A}$ of an amplifier used to control the TEM. This amplifier is controlled in high-gain feedback, therefore it adjusts its output voltage to compensate for the $V_{S_m}$ that acts as a back EMF type voltage. This voltage $V_{S_m}$ in $\mathrm{V}$ is known as the Seebeck effect, and it generates a voltage based on the temperature gradient over the TEM.
\paragraph*{Time constants} Solving \eqref{eq:tem_voltage} for 2 unknowns is generally not possible. However, as suggested in \citet{mitrani_methodology_2004}, $V_{R_m}$ and $ V_{S_m}$ manifest in different time scales. This difference in time constants is illustrated in Fig. \ref{fig:peltiervoltage}. At time $t_a$, a current command $\delta I$ is applied to the amplifier, causing an instantaneous step in electric potential $V_{R_m}$. While $V_{S_m}$ only manifests after a sufficient time has passed and a thermal equilibrium is reached at time $t_b$ yielding a $\Delta T = T_h - T_c$ over the TEM. By explicitly exploiting this difference in time constants, both $S_m(T_{avg})$ and $R_m(T_{avg})$ can be determined from \eqref{eq:tem_voltage} using voltage measurements. 
\section{Experimental identification}\label{sec:5}
In this section the temperature dependent parameters are identified using the approach presented in Sec. \ref{sec3_2}. The parameters are identified using a dedicated experimental setup.
\subsection{Experimental Identification setup}
\begin{figure}
	\centering
	\includegraphics[width=1\linewidth]{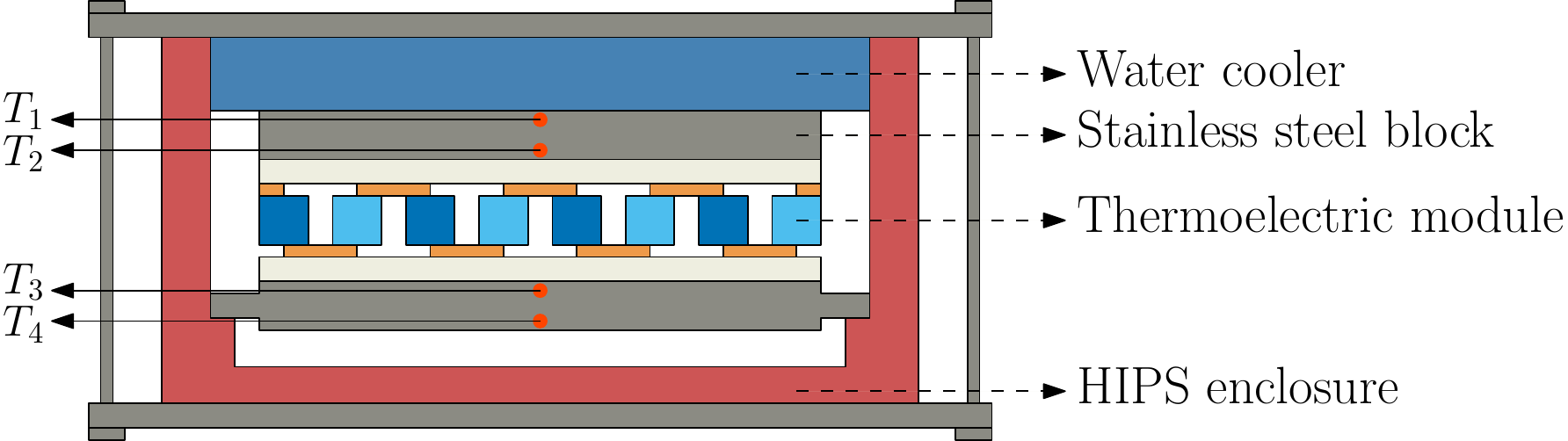}
	\caption{Schematic representation of the experimental setup, its various components and sensor locations.}
	\label{fig:peltier_setup}
\end{figure}

\begin{figure}
	\centering
	\setlength{\fboxsep}{0pt}
	\fbox{\includegraphics[width=0.48\linewidth]{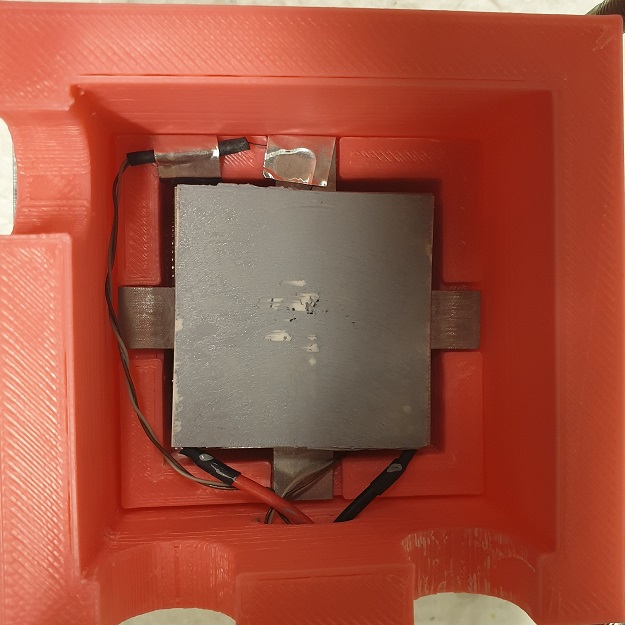}}
	\fbox{\includegraphics[width=0.48\linewidth]{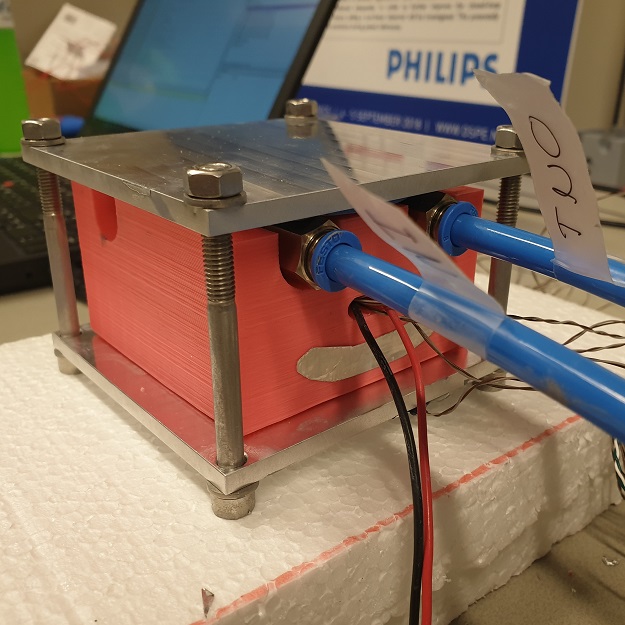}}
	\caption{Photographs of the experimental identification setup. On the left the internal compartment is shown, illustrating a small thermal connection to the HIPS enclosure by reducing the contact area of the support. On the right, the full setup is shown including the connections to the water chiller that provides conditioned water to the heatsink.}
	\label{fig:exp_setup}
\end{figure}

A dedicated TEM parameter identification setup is designed to isolate the TEM from external influences and facilitate accurate estimation of temperature dependent parameters.
In Fig. \ref{fig:peltier_setup} a schematic representation of the setup is shown. The TEM is clamped between two stainless steel blocks to provide some additional thermal mass and spread the heat evenly. On the top, the hot side, the steel block is conditioned using a water cooling block and water chiller to provide a temperature stable heat sink. The setup is encapsulated by a 3D-printed enclosure made of High Impact PolyStyrene (HIPS) that is printed with a low infill of 10$\%$ to provide thermal insulation from the environment. 
The temperature measurements are done using thermistors with negative temperature coefficients, or NTC for short. Each stainless steel block contains 2 NTC sensors, as indicated in Fig. \ref{fig:peltier_setup}, where $T_2$ and $T_3$ are considered the TEM hot and cold side respectively. To mitigate heat transfer from the setup to the enclosure, small tabs connect the lower block to the HIPS enclosure, as shown in Fig. \ref{fig:exp_setup}, to minimize the contact area.
\paragraph*{Data acquisition} To measure the temperature, voltages and current in the TEM identification setup a CompactDAQ by National Instruments is used. To facilitate temperature measures using the NTC sensors, a Wheatstone bridge is used that converts the resistance measurements, and thereby the temperature, to an electrical potential. Moreover, since the identification method proposed in Sec. \ref{sec:method_id_param} relies heavily on the known input current, a precision power resistor is placed in series with the TEM. The resistor is selected such that its resistance remains constant for the operating currents. By measuring the voltage drop over the resistor the current can be accurately calculated. 
\subsection{Temperature dependent identification} \label{sec:case_temp_param}
\begin{figure*}
	\centering
	\pgfplotsset{every axis plot/.append style={line width=1pt}}
	\begin{subfigure}{0.33\linewidth}
		\includegraphics[width=1\linewidth]{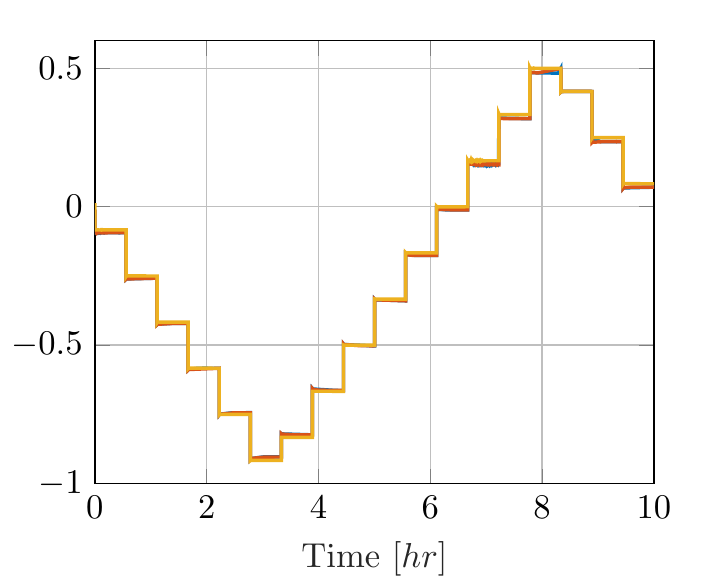}
		\caption{Current $I_{TEM}$ in $\mathrm{A}$}
		\label{fig:sim_transient_a}
	\end{subfigure}%
	\begin{subfigure}{0.33\linewidth}
		\includegraphics[width=1\linewidth]{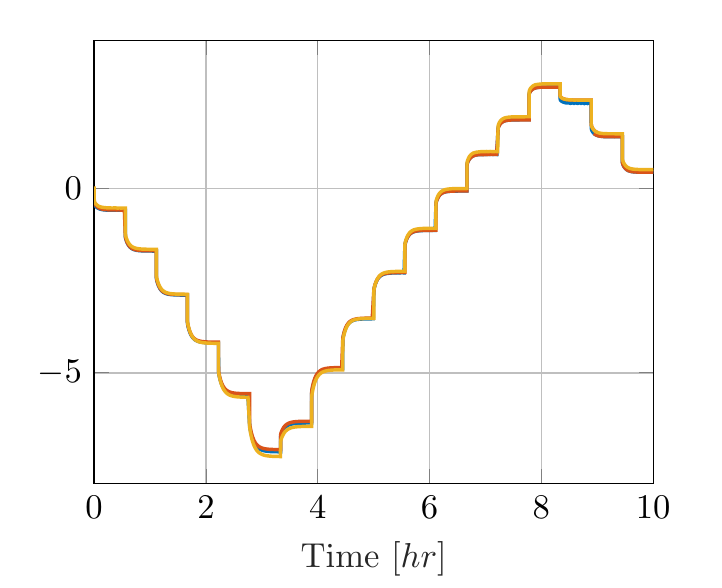}
		\caption{Voltage $V_{TEM}$ in $\mathrm{V}$}
		\label{fig:sim_transient_b}
	\end{subfigure}%
	\begin{subfigure}{0.33\linewidth}
		\includegraphics[width=1\linewidth]{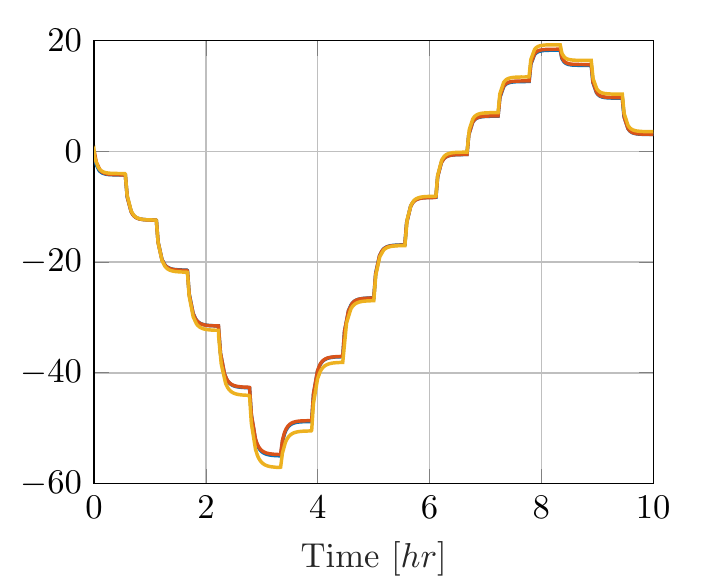}
		\caption{$\Delta$ T in $\mathrm{^\circ C}$}
		\label{fig:sim_transient_c}
	\end{subfigure}
	\caption{Identification experiment used to identify the temperature dependent electrical resistance $R_m(T)$ and Seebeck coefficient $S_m(T)$. The different sub-plots show the Current, Voltage and Temperature respectively. The experiment is repeated for $3$ modules, TEM 1 \tikzline{mycolor1}, TEM 2 \tikzline{mycolor2} and TEM 3 \tikzline{mycolor3}. }
	\label{fig:sim_transient}
\end{figure*}

\begin{figure}
	\centering
	\includegraphics[]{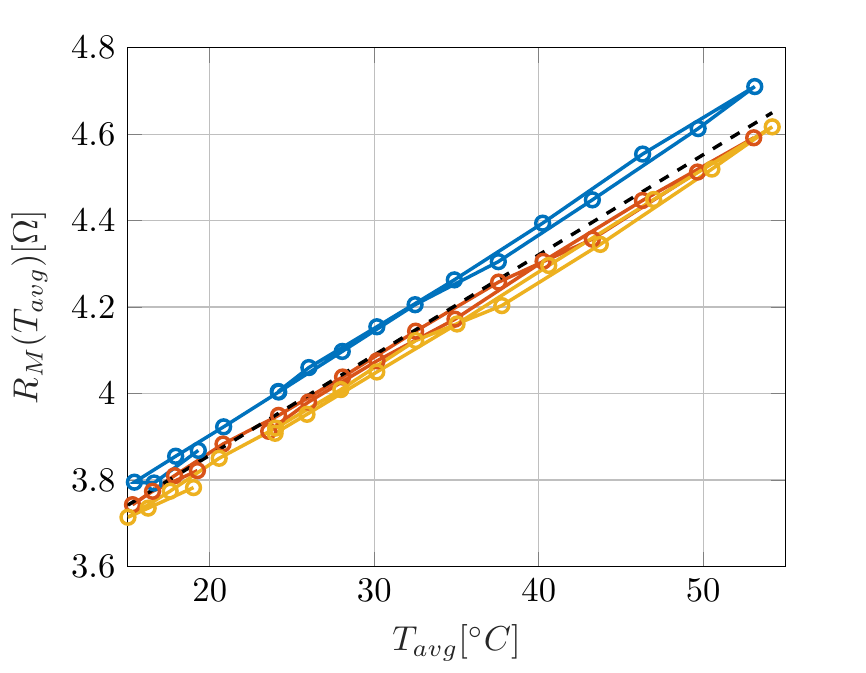}
	\caption{Identifying the temperature dependent electrical resistance $R_m(T)$ for different peltier modules. It shows that for TEM 1 \tikzmarkline{mycolor1}, TEM 2 \tikzmarkline{mycolor2} and TEM 3 \tikzmarkline{mycolor3} the results show a similar linear relation with the average temperature $T_{avg}$ for all TEMs leading to an average $R_m(T_{avg})$ \tikzdashedline{black}.}
	\label{fig:T_Dep_Param_Rm}
\end{figure}

\begin{figure}
	\centering
	\includegraphics[]{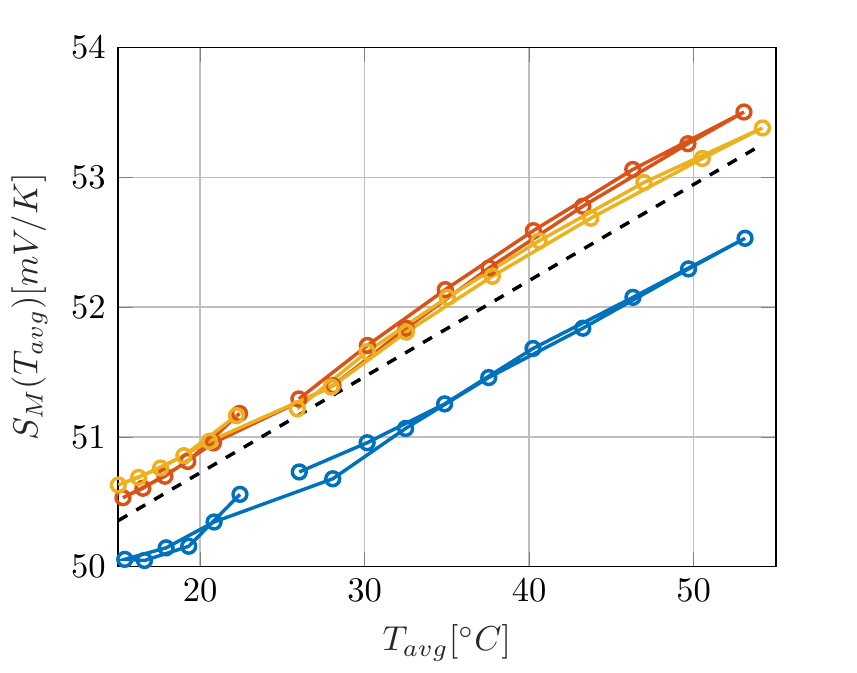}
	\caption{Identifying the temperature dependent Seebeck coefficient $S_m(T)$ for different TEMs. It shows that for TEM 1 \tikzmarkline{mycolor1} and TEM 2 \tikzmarkline{mycolor2} the result is quite similar, and TEM 3 \tikzmarkline{mycolor3} deviates from the rest. This yields a slightly shifted average linear relation for $S_m(T_{avg})$ \tikzdashedline{black}.}
	\label{fig:T_Dep_Param_Sm}
\end{figure}

In this section the method proposed in Sec. \ref{sec:method_id_param} is utilized to estimate the temperature dependent parameters $R_m(T_{avg}), S_m(T_{avg})$ for multiple TEMs. 
To yield an accurate model for the purposes of this paper, a significant temperature range for $T_{avg}$ must be considered. To achieve this, the input current $I$ is changed in small increments covering a wide operating range, as shown in Fig. \ref{fig:sim_transient_a}. 
\paragraph*{Identification procedure}
The identification procedure of the temperature dependent parameters can be described as
\begin{center}
	\begin{algorithm}
		\caption{Identification procedure}
		\label{alg:parameter}
		\begin{algorithmic}
			\State Initalize $I = I_0$
			\For{Each $\delta I$}
			\State I + $\delta I$
			\State Calculate $R_m(T_{avg}) = \delta I/V_{R_m}$
			\Ensure Steady-State
			\State  Calculate $S_m(T_{avg}) = V_{S_m}/\Delta T$
			\EndFor
		\end{algorithmic}
	\end{algorithm}
\end{center}
where $\delta I$ are the steps in the current reference for the amplifiers, as shown in Fig. \ref{fig:sim_transient_a} and $I_0 = 0$ is the initial current. The electrical resistance $R_m(T_{avg})$ is estimated from the instantaneous voltage jump $V_{R_m}$, shown in Fig. \ref{fig:sim_transient_b}, that occurs after a step in current, since $R_m(T_{avg}) = I/V_{R_m}$. Then, the current is maintained until the system reaches a steady-state and associated $\Delta T$, as shown in \ref{fig:sim_transient_c}. This yields a back EMF voltage due to the Seebeck effect $V_{S_m} = S_m(T_{avg})\Delta T$, that is used to estimate $S_m(T_{avg})$. By repeating this process both parameters are estimated for a range of $T_{avg}$.
\paragraph*{Temperature dependent parameters}
The identification procedure is repeated for 3 different, but of equal type, TEMs. This allows the characterization of an average parameter over a batch of actuators. While individual calibration curves could yield superior results, most applications do not allow for dedicated unit calibration tests, since they are both time intensive and expensive. The results of the Identification procedure are shown in Fig. \ref{fig:T_Dep_Param_Rm} for $R_m(T_{avg})$ and in Fig. \ref{fig:T_Dep_Param_Sm} for $S_m(T_{avg})$. Both parameters show a linear dependency on $T_{avg}$ and a significant change of their value over the range of $15^\circ C$ to $55^\circ C$. The results show a small spread over the different TEMs. However, in Fig. \ref{fig:T_Dep_Param_Sm} the first module has a slightly different $S_m(T_{avg})$, this could be an outlier but a larger sample size is required to yield a more definitive outcome. Moreover, since the input current profile consists of both positive and negative $\delta I$ steps, at similar $T_{avg}$, some insight into possible hysteresis effects is gained. In the results, the positive and negative current perturbation yield similar parameter estimates, indicating that hysteresis effects are negligible.
\section{Experimental validation}\label{sec:6}
In this section the temperature dependent parameters are included in the TEM setup model to achieve improved simulation results. The model is compared to experimental measurements on the setup.
\subsection{Model}
To validate the effectiveness of the procedure proposed in Sec. \ref{sec:method_id_param} and the results obtained in Sec. \ref{sec:case_temp_param} a full thermodynamical model of the setup shown in Fig. \ref{fig:exp_setup} is constructed. The model is obtained in state-space form, similar to \eqref{eq:state_space}, and the thermoelectric dynamics and temperature dependent parameters are included in the nonlinear contribution $F^{NL}(x,u)$. The remainder of the model consists of a lumped representation of the TEM, stainless steel blocks, HIPS enclosure and water cooler, see Fig. \ref{fig:peltier_setup}. The model parameters, e.g., the resistances and thermal capacitances, are optimized using a nonlinear databased optimization procedure, seeded with initial values based on first-principles. 
\begin{figure}
	\centering
	\includegraphics[width=0.95\linewidth]{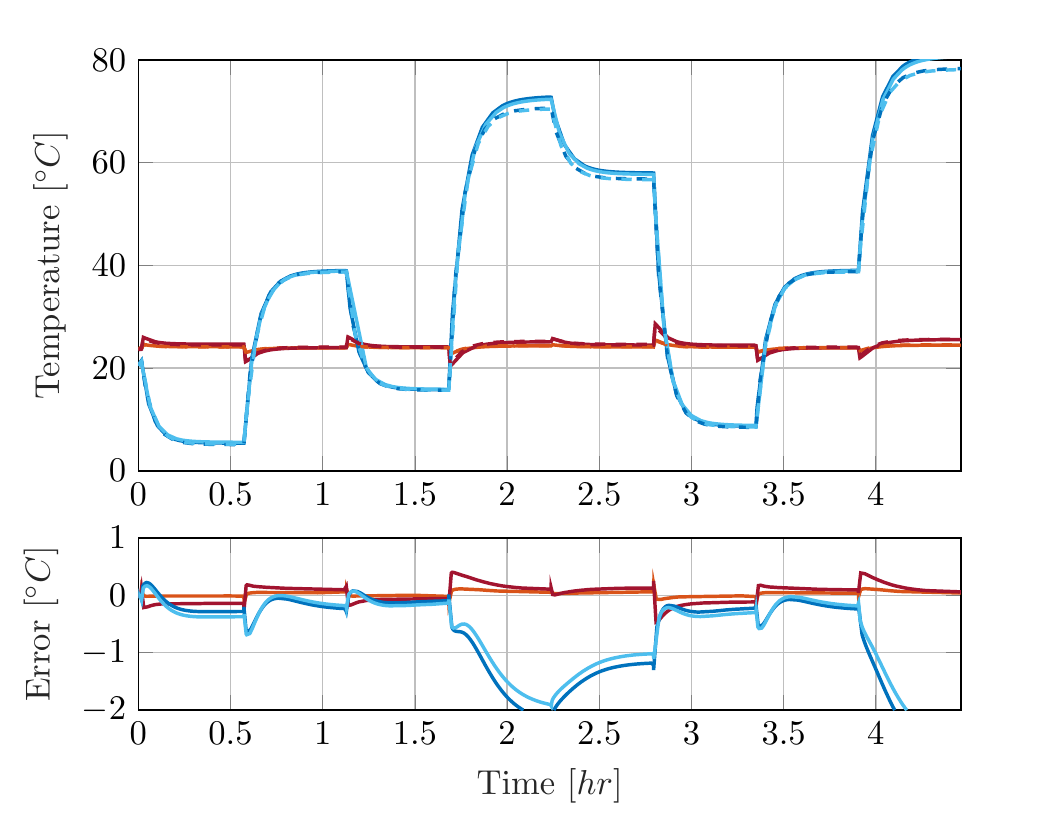}
	\caption{Simulation results (dashed) compared to experimental measurements (solid) using constant parameters at $T_{avg} = 35^\circ C$ for $T_1$ \tikzline{mycolor2}, $T_2$ \tikzline{mycolor5}, $T_3$ \tikzline{mycolor1}, $T_4$ \tikzline{mycolor6}. 
		The results show that at temperatures significantly different from $T_{avg} = 35^\circ C$ the model is inaccurate since temperature dependency must be taken into account. }
	\label{fig:Sim_Peltier_P2_Tavg_35}
\end{figure}
\subsection{Validation: Constant parameters}
To verify the improved estimation accuracy of the model by including the temperature dependency in the parameters, a validation dataset is employed. The parameters of the model are optimized using the identification dataset shown in Fig. \ref{fig:sim_transient}. The parameters $S_m$ and $R_m$ are fixed at their values at $T_{avg} = 35^\circ C$, that is considered an average temperature in the experiment. The resulting model is then used to yield simulation results as shown in Fig. \ref{fig:Sim_Peltier_P2_Tavg_35}. 
It shows that the model is not able to capture accurately the system dynamics at temperatures other than the assumed $T_{avg} = 35^\circ C$. 
\subsection{Validation: Improved accuracy}
\begin{figure}
	\centering
	\includegraphics[width=0.95\linewidth]{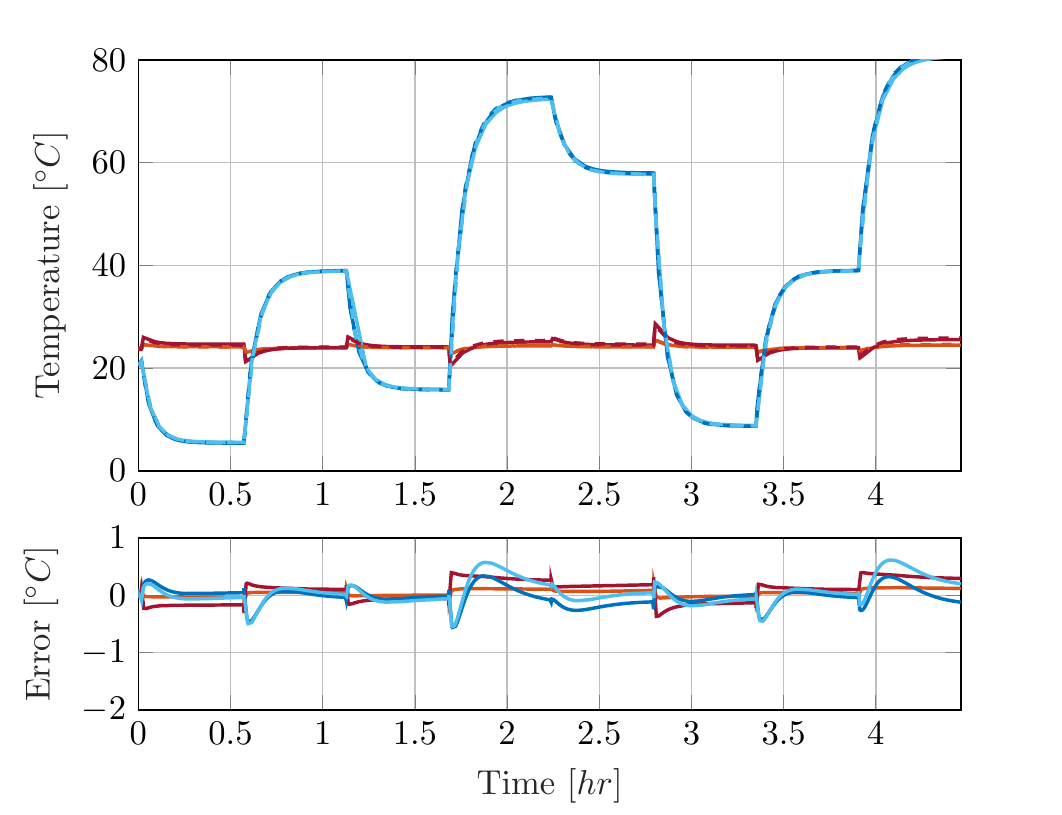}
	\caption{Simulation results (dashed) compared to experimental measurements (solid) using temperature dependent parameters for $T_1$ \tikzline{mycolor2}, $T_2$ \tikzline{mycolor5}, $T_3$ \tikzline{mycolor1}, $T_4$ \tikzline{mycolor6}. The model prediction error is significantly improved to results in Fig. \ref{fig:Sim_Peltier_P2_Tavg_35} by taking into account temperature dependent parameters. }
	\label{fig:Sim_Peltier_P2_Tavg_var}
\end{figure}

By including the temperature dependent parameters in the simulation model the simulation error can be reduced. The temperature dependent parameters $R_m(T_{avg}),S_m(T_{avg})$ are now included as linear relations in the model. Results shown in Fig. \ref{fig:Sim_Peltier_P2_Tavg_var} illustrate that by taking into account the temperature dependent parameters a significant improvement in model accuracy is achieved. The model error residual now shows little correlation to the operating temperature, illustrating that the model now more accurately captures the thermal dynamics of the system for a wide range of operating temperatures. 
It is expected that a model with increased complexity could further reduce the prediction error of the temperature. However, for the intended application, the simplified model that yields an error of $ \pm 0.5 ^\circ C$ is sufficiently accurate.
\section{Conclusion}
In this work, it is shown that TEMs can potentially yield significant benefits in active thermal control for various industrial applications. To facilitate advanced control approaches and to achieve accurate temperature observers a high-fidelity model is required. To yield sufficient model accuracy, temperature dependency of the model parameters must be taken into account. By applying the approach presented in this work, these temperature dependent parameters are incorporated into the thermodynamical model. The procedure yields a high-fidelity model that is accurate over a wide range of operating temperatures.


\end{document}